\documentclass[pra,aps,amsmath,amssymb,amsfonts,twocolumn,nofootinbib,longbibliography,superscriptaddress]{revtex4-2}
\usepackage{amssymb,diagbox,multirow}
\usepackage{bm,mathrsfs}
\usepackage{graphicx}
\usepackage{epsfig}
\usepackage{amsmath,bbm}
\usepackage{amsfonts,amssymb}
\usepackage{times}
\usepackage{verbatim}
\usepackage[sort&compress]{natbib}
\usepackage{amsmath}
\usepackage[colorlinks=true,citecolor=blue,linkcolor=blue,urlcolor=blue]{hyperref}
\usepackage[usenames]{color}

\newcommand{\bra}[1]{\langle #1|}
\newcommand{\ket}[1]{|#1\rangle}

\begin{document}

\title{Ground-state chiral current via periodic modulation}

\author{Shuyue Wang}
\affiliation{Center for Quantum Sciences and School of Physics, Northeast Normal University, Changchun 130024, China}

\author{Wuji Zhang}
\affiliation{Center for Quantum Sciences and School of Physics, Northeast Normal University, Changchun 130024, China}

\author{Chunfang Sun}
\affiliation{Center for Quantum Sciences and School of Physics, Northeast Normal University, Changchun 130024, China}

\author{Chunfeng Wu}
\affiliation{Science, Mathematics and Technology, Singapore University of Technology and Design, 8 Somapah Road, Singapore 487372, Singapore}


\author{Gangcheng Wang}
\email{wanggc887@nenu.edu.cn}
\affiliation{Center for Quantum Sciences and School of Physics, Northeast Normal University, Changchun 130024, China}
\affiliation{Key Lab of Advanced Optical Manufacturing Technologies of Jiangsu Province, Soochow University, Suzhou 215006, China}
\date{\today}

\begin{abstract}

In this study, we engineer the $z$-component of Dzyaloshinskii-Moriya interaction mediated by photons to emulate ground-state chiral current based on three-level atoms driven by quantum and classical fields. We employ adiabatic elimination techniques to derive an effective Dzyaloshinskii-Moriya interaction Hamiltonian of two-level systems in the ground-state manifold, which can address the challenges arising from the finite lifetime of excited states. Furthermore, we can ensure to achieve desired dynamics through the implementation of periodic modulation on the atomic ground states. Meanwhile, three-state and multi-state chiral current can be obtained by choosing appropriate driving frequencies and phases. The Dzyaloshinskii-Moriya interaction for other components can also be designed based on a toggling frame. The numerical simulation results further indicate that our proposal can generate the perfect three-state chiral current and reliable multi-state chiral current. Our results provide possibilities for quantum state transfer and future quantum networks.

\end{abstract}

\maketitle

\section{Introduction}
\label{SecI}

Photon-mediated excitation exchange \cite{PhysRevLett.88.243602,PhysRevA.66.022314,PhysRevA.81.021804}, as a well-established paradigm in quantum simulations, has garnered significant attention due to its pivotal role in testing the fundamentals of quantum mechanics \cite{Chiocchetta2021,PhysRevLett.114.173002}. In particular, spin-exchange interactions mediated by vacuum fields within cavities have been investigated from solely scientific discovery to the foundational engineering associated with the design of multi-qubit quantum systems \cite{doi:10.1126/science.aar3102,PhysRevA.96.050301,PhysRevLett.121.070403}. Overall, the optical manipulation of analogous interactions involves the two-photon coupling between energy levels including hyperfine and Zeeman states. Such approach allows for the encoding of spin states in long-lived ground states, which can enable to generate remarkable coherent characteristics and robust exchangeable scheme \cite{Morgado_2021,PhysRevLett.123.170503,Picken_2019,Jau_2015,PhysRevLett.123.230501}.
On the other hand, this encoding provides the intriguing possibility for further exploring higher spin models \cite{PhysRevLett.119.213601,PhysRevLett.104.195303,doi:10.1126/science.aag1106} and the preparation of squeezed states quantum metrology \cite{PhysRevLett.104.073602,doi:10.1126/science.aaf3397}. 

Dzyaloshinskii-Moriya interaction (DMI) \cite{PhysRevLett.4.228,PhysRev.120.91}, originating from relativistic antisymmetric exchange interactions, can induce a range of chiral phenomena, such as spin spirals and skyrmions \cite{RevModPhys.89.025006,Haze2017}, which have been harnessed in spintronics to engineer spin structures with topological chiral conservation. Recently, DMI has been simulated based on Floquet engineering \cite{PhysRevA.108.053318} which offers a route for exploring quantum spin chirality. Meanwhile, spin chirality originating from antisymmetric spin exchange interactions has been experimentally verified in the superconducting qubits platform \cite{Wang2019}. Unlike the spin chiral operators \cite{Liu2020SynthesizingTI}, the $z$-component of DMI breaks parity symmetry (exchanging $\vec{\sigma}_{j}$ with $\vec{\sigma}_{k}$) while preserving time-reversal symmetry (replacing $\vec{\sigma}_{j}$ by $-\vec{\sigma}_{j}$). This property enables the realization of both the chiral spin current and two opposing directions of spin-excited chiral evolution while reversing the initial states of all spins \cite{Wang2019}. It is worthy that the phenomenon is similar to the quantum spin Hall effect \cite{PhysRevLett.103.047203,PhysRevLett.115.106603,PhysRevB.99.054409,PhysRevB.100.064429}, where electrons traveling in opposite directions along the edges of a material possess opposing spins. Up to now, spin chirality \cite{PhysRevB.39.11413} has been successfully implemented on various platforms, such as hybrid cavity-magnon systems \cite{PhysRevA.106.033711}, synthetic magnetic fields \cite{PhysRevLett.126.103603}, superconducting circuits \cite{Roushan2017}, optical tweezer arrays \cite{PhysRevA.105.032417}, Rydberg atoms \cite{PhysRevResearch.4.L032046,kuznetsova2023engineering}, circuit QED \cite{PhysRevB.87.134504,PhysRevB.103.224525,PhysRevA.108.063709}, and ion clusters \cite{PhysRevLett.123.213605}. Due to its favorable properties, it holds promise in the preparation of entangled states \cite{Wang2019}, the quantum Hall effect \cite{doi:10.1126/science.1058161}, chiral spin liquids \cite{Bauer2014,RevModPhys.89.025003}, quantum state transfer \cite{wang2023chiral}, chiral separation \cite{Tao2021} and stabilization of skyrmions \cite{2013NatNa...8..899N,Tokunaga2015}. Nevertheless, these previous studies have demonstrated perfect chiral current in a triangular three-node network and failed to work in larger networks. The existence of flawless chiral current in multi-node networks remains an open question, which is crucial for advancing applications in lattices with non- triangular structure.

In parallel, periodic modulation \cite{PhysRevX.4.031027, PhysRev.138.B979, PhysRevB.97.125429, 10.1063/5.0033382, PhysRevA.109.013712} stands as a pivotal technique across signal processing, communications, and control systems, characterized by the transmission of information through systematic alterations in specific signal parameters. This methodology encompasses a spectrum of techniques such as amplitude modulation \cite{PhysRevA.105.012203}, frequency modulation \cite{Silveri_2017}, and phase modulation \cite{ZHANG2015183}, among others. These approaches significantly enhance the controllability of quantum systems and pave the way for the development of more robust schemes. An intriguing question arises: can the incorporation of periodic modulation into ground-state chiral current yield analogous beneficial outcomes? It motivates us to delve into the exploration of chiral current through periodic modulation.

In this paper, we propose a theoretical scheme for realizing DMI within the hyperfine ground-state manifold of three-level atoms by applying periodic modulation,
which gives rise to the generation of perfect chiral current in three-level atomic chains. A series of transformations have been applied to the Hamiltonian to derive the effective interaction. Specifically, we investigate that how to generate three- and multi- state chiral current with opposing directions in the practical system. Additionally, by utilizing the toggling frame \cite{PRXQuantum.4.010334}, we can obtain three components of DMI from the $z$ component of the effective Hamiltonian. Furthermore, the accurate description of high-fidelity in the numerical simulation results reveal that our scheme can achieve extremely excellent chiral current in the ground state.

The remainder of this paper is organized as follows: In Sec. \ref{SecII}, we present a detailed process for the derivation of the effective Hamiltonian and discuss the dynamics of three- and multi-state chiral current by adjusting the phase and driving frequencies. Subsequently, we construct pulse sequences on the $z$ component to derive other components of DMI. In Sec. \ref{SecIII}, we analyze the influence of decoherence on the ground-state chiral current and consider experimentally feasible parameters. The simulation results show that our scheme can generate desired ground-state chiral current. Finally, we provide a comprehensive summary of the entire paper in Sec. \ref{SecV}.

\section{The effective DMI and chirality}
\label{SecII}

\subsection{The derivation of the effective Hamiltonian}
\label{SecIIA}

\begin{figure}[htbp]
    \centering
    \includegraphics[scale=0.36]{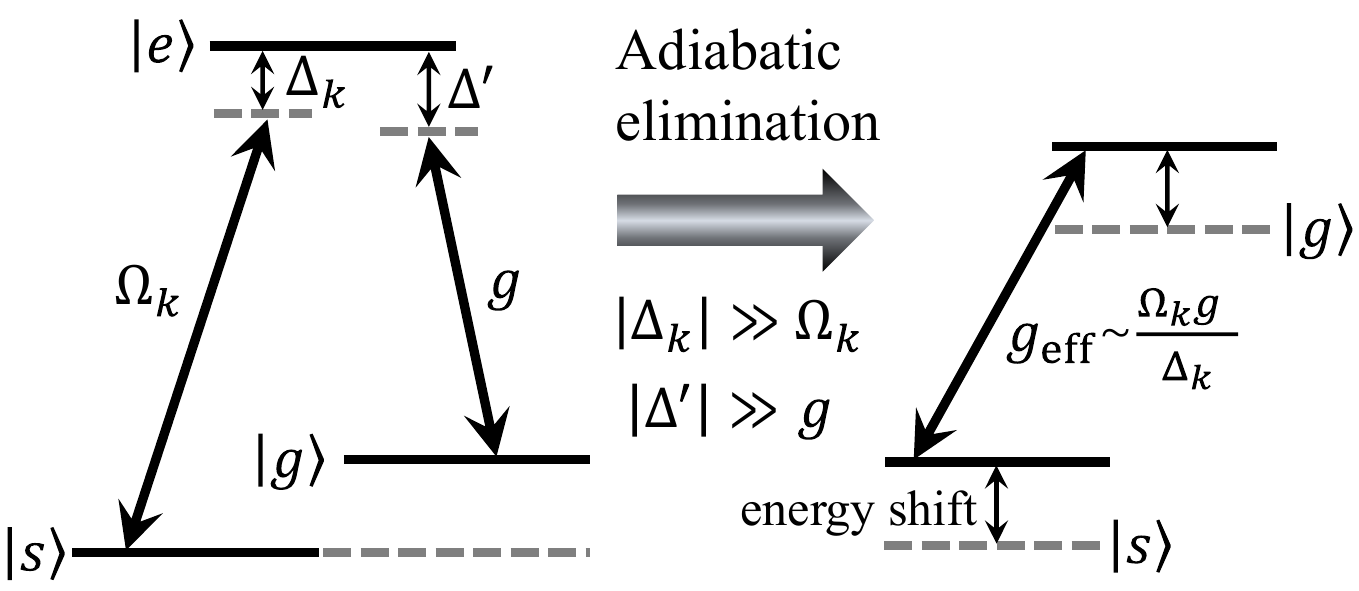}
    \caption{Schematic diagram of the $\Lambda$-type three-level atom. $\{\ket{s}, \ket{g}, \ket{e}\}$ represent three atomic energy levels with energy $\{0, \omega_{g}, \omega_{e}\}$, respectively. The pump laser with Rabi frequency $\Omega_{k}$ and frequency $\omega_{k}^{L}$ induces $\sigma$ transition between atomic ground $\ket{s}$ and excited state $\ket{e}$. A photon mode with coupling strength $g$ and frequency $\omega_{c}$ induces $\pi$-transition between $\ket{g}$ and $\ket{e}$. Under the condition of large detuning (i.e., $|\Delta_{k}| \gg \Omega_{k}$ and $|\Delta'| \gg g$), the transitions between the two ground states $\ket{s}$ and $\ket{g}$ with the effective coupling strength
    $g_{\rm eff} \sim \frac{\Omega_{k}g}{\Delta_{k}}$ can be mediated by adiabatic elimination.}
    
    \label{fig01}
\end{figure}

We consider to simulate DMI and chiral current with trapping cold atoms ($\rm ^{87}Rb$) in an intracavity lattice \cite{PhysRevResearch.4.013089,PhysRevLett.125.060402,PhysRevLett.122.010405}. In our proposed scheme, the common single-mode cavity mediates the interactions between atoms at different lattice sites. In Fig. \ref{fig01}, we adopt a three-level system in a $\Lambda$ configuration, consisting of two lower energy states $\ket{s}$ and $\ket{g}$, along with an excited state $\ket{e}$.
The pump beam and the cavity mode separately couple two legs of the configuration, while applying the periodic driving field at the atomic ground-state transition frequency. The Hamiltonian is initially written under the rotating wave approximation as (hereafter we set $\hbar = 1$) 
\begin{equation}\label{eq_01}
  \hat{H}(t)=\hat{H}_{C}+\hat{H}_{A}(t)+\hat{H}_{L M}+\hat{H}_{P}(t),
\end{equation}
where
\begin{subequations}\label{eq_02}
  \begin{align}
     \hat{H}_{C} & =\omega_{c} \hat{a}^{\dagger} \hat{a}, \label{eq_02a} \\  
     \hat{H}_{A}(t) & =\sum_{k}\omega_{e} \hat{\sigma}_{k}^{ee}+\omega_{k}^{g}(t) \hat{\sigma}_{k}^{gg}, \label{eq_02b} \\  
     \hat{H}_{L M} & =\sum_{k} g \hat{a} \hat{\sigma}_{k}^{eg}+{\rm H.c.}, \label{eq_02c} \\  
     \hat{H}_{P}(t) & =\sum_{k} \frac{\Omega_{k}}{2} \hat{\sigma}_{k}^{se} e^{i \omega_{k}^{L} t}+{\rm H.c.}. \label{eq_02d}
  \end{align}
\end{subequations}
The notation $\hat{a}$ ($\hat{a}^{\dagger}$) is the annihilation (creation) operator of the cavity field with frequency $\omega_{c}$. The notation $\omega_{e}$ represents the assisted excited state $\ket{e}$ of the atomic energy and the two internal atomic states $\{\ket{s}, \ket{g}\}$ in the hyperfine manifold denote the pseudospin-1/2 state with the energy level splitting $\omega_{k}^{g}(t)$, where $\hat{\sigma}_{k}^{ab} = \ket{a}_{k}\bra{b}$ corresponds to three atomic energy levels $a, b = \{s, g, e\}$ and $k$ marks the position. The periodic driving field takes the form $\omega_{k}^{g}(t) =\omega_{g}+\varepsilon_{k}\cos(\nu_{k}t-\phi_{k})$ with the amplitude $\varepsilon_{k}$, frequency $\nu_{k}$ and phase $\phi_{k}$ acting on $k$th atom, respectively. The notation $g$ denotes the coupling strength between atoms and cavity. While the transition between $\ket{s}$ and $\ket{e}$ is driven by an external local pump field with frequency $\omega_{k}^{L}$ and Rabi frequency $\Omega_{k}$.

Moving to the rotation frame defined by
$\hat{U}(t)=e^{-i\sum_{k}(\omega_{k}^{L}\hat{\sigma}_{k}^{ee}+\omega_{g}\hat{\sigma}_{k}^{gg})t}e^{-i\omega_{c}\hat{a}^{\dagger}\hat{a}t}$, the transformed Hamiltonian can be expressed as
\begin{equation}\label{eq_03}
  \begin{split}
    \hat{\tilde{H}}(t) & =\hat{U}^{\dagger}(t)\left[\hat{H}(t)-i \partial_{t}\right] \hat{U}(t) \\
    & = \hat{\tilde{H}}_{A}(t)+\hat{\tilde{H}}_{L M}(t)+\hat{\tilde{H}}_{P},
  \end{split}
\end{equation}
where
\begin{subequations}\label{eq_04}
  \begin{align}
   \hat{\tilde{H}}_{A}(t) & =\sum_{k}\Delta_{k} \hat{\sigma}_{k}^{ee}+\varepsilon_{k}\cos(\nu_{k}t-\phi_{k}) \hat{\sigma}_{k}^{gg}, \\
   \hat{\tilde{H}}_{LM}(t) & =\sum_{k} g \hat{a} \hat{\sigma}_{k}^{eg}e^{i (\Delta'-\Delta_{k}) t}+{\rm H.c.}, \\
   \hat{\tilde{H}}_{P} & =\sum_{k} \frac{\Omega_{k}}{2} \hat{\sigma}_{k}^{se}+{\rm H.c.}.
  \end{align}
\end{subequations}
Here $\Delta_{k}=\omega_{e}-\omega_{k}^{L}$ and $\Delta'=\omega_{e}-\omega_{g}-\omega_{c}$. In the large detuning framework, the frequencies of both the cavity mode and Raman fields are far detuned to the atomic transition, i.e., $|\Delta_{k}| \gg \Omega_{k}$ and $|\Delta'| \gg g$. Assuming a $\Lambda$-type system characterized by a significant detuning between atoms and driving fields, the technique of adiabatic elimination \cite{doi:10.1073/pnas.1603777113, Brion_2007, gonzalezballestero2023tutorial} can be employed to derive an effective Hamiltonian. For the system considered in this work, we can introduce two orthogonal projection operators $\hat{P}=\sum_{k}(\hat{\sigma}_{k}^{gg}+\hat{\sigma}_{k}^{ss}), \hat{Q}=\sum_{k}\hat{\sigma}_{k}^{ee}$. The effective Hamiltonian can be recast as  (more details are presented in Appendix \ref{appendixA})
\begin{equation}\label{eq_06}
 \begin{split}
    \hat{H}'(t) = & \hat{P} \hat{\tilde{H}}(t) \hat{P}-\hat{P} \hat{\tilde{H}}(t) \hat{Q} \frac{1}{\hat{Q} \hat{\tilde{H}}(t) \hat{Q}} \hat{Q} \hat{\tilde{H}}(t) \hat{P} \\
    =& \hat{H}'_{0}(t)+\hat{H}'_{\rm int}(t),
  \end{split}
\end{equation}
where
\begin{subequations}
    \begin{align}
        \hat{H}'_{0}(t)=& \sum_{k} \left[\omega_{k}^{g}(t)-\omega_{g}-\frac{g^{2}}{\Delta_{k}} \hat{a}^{\dagger} \hat{a}\right]\hat{\sigma}_{k}^{gg}-\sum_{k} \frac{\Omega_{k}^{2}}{4 \Delta_{k}} \hat{\sigma}_{k}^{ss}, \\
        \hat{H}'_{\rm int}(t)=& -\sum_{k}\left[\frac{\Omega_{k} g}{2 \Delta_{k}} \hat{a}^{\dagger} \hat{\sigma}_{k}^{gs}e^{-i(\Delta'-\Delta_{k})t}+{\rm H.c.}\right].
    \end{align}
\end{subequations}
Following the adiabatic elimination of the excited state, each atom can be considered as a $gg$-type qubit. Due to the large detuning in frequency between the photon mode and the atom, it is obvious that the excitation process can not be achieved through photon-atom interaction and the mode $a$ will not be excited \cite{PhysRevResearch.4.033083}. Therefore, the term $\sum_{k}(g^{2}/\Delta_{k})\hat{a}^{\dagger}\hat{a}\hat{\sigma}_{k}^{gg}$, which scales with the average photon number, can be safely neglected.

For the sake of facilitating the discussion, the Hamiltonian is converted into the interaction picture with respect to the following time-dependent unitary operator
\begin{equation}\label{eq_07}
\small
     \begin{split}
     \hat{U}_{0}(t) & =\exp \left[-i \int_{0}^{t} dt' \hat{H}'_{0}\left(t'\right)\right] \\
     & =\exp \left\{-i \sum_{k} \left[\frac{\varepsilon}{\nu} \sin (\nu t-\phi_{k})+\beta_{k}\right]\hat{\sigma}_{k}^{gg} + i \sum_{k}\frac{\Omega_{k}^{2}}{4 \Delta_{k}} \hat{\sigma}_{k}^{ss} t\right\},
    \end{split}
\end{equation}
where $\beta_{k}=(\varepsilon/\nu) \sin \phi_{k}$. For convenience, we assume the uniform amplitude ($\varepsilon_{k} = \varepsilon$) and frequency ($\nu_{k} = \nu$) here. One obtains the interaction between photons and qubits with
$\delta_{k}=\Delta'-\Delta_{k}-\Omega_{k}^{2}/4\Delta_{k}$, and the following transformed Hamiltonian
\begin{equation}\label{eq_08}
   \begin{split}
    \hat{H}_{I}(t) & =\hat{U}_{0}^{\dagger}(t)\left[\hat{H}^{\prime}(t)-i \partial_{t}\right] \hat{U}_{0}(t) \\
    & =-\sum_{k} \frac{\Omega_{k} g}{2 \Delta_{k}} \hat{a} \hat{\sigma}_{k}^{sg}e^{i\left\{\delta_{k} t-\left[\frac{\varepsilon}{\nu} \sin (\nu t-\phi_{k})+\beta_{k}\right]\right\}}+{\rm H.c.}.
   \end{split}
\end{equation}
It is worthy mentioning that the site-dependent phase factor $\beta_{k}$ can be gauged away from the Hamiltonian by performing the unitary transformation $\hat{R}=\prod_{k=1}^{N}e^{i\beta_{k}\hat{\sigma}_{k}^{z}/2}$ on the Hamiltonian in Eq. (\ref{eq_08}). Then we can tune $\Delta_{k}$ and $\Delta'$, so that $\delta_{k}=0$ (i.e., $\Delta'=\Delta_{k}+\Omega_{k}^{2}/4\Delta_{k}$). Using the Jacobi-Anger identity $e^{i z\sin\theta}=\sum_{n=-\infty}^{\infty}\mathcal{J}_{n}(z)e^{i n\theta}$ where $\mathcal{J}_{n}(z)$ represents the $n$-th order Bessel function of the first kind, the Hamiltonian in Eq. (\ref{eq_08}) can be rewritten as
\begin{equation}\label{eq_09}
\small
    \hat{H}_{I}(t)=-\sum_{k} \sum_{n=-\infty}^{\infty} \frac{\Omega_{k} g}{2 \Delta_{k}} \mathcal{J}_{n}\left(-\frac{\varepsilon}{\nu}\right) e^{i n\left(\nu t-\phi_{k}\right)}\hat{a} \hat{\sigma}_{k}^{sg}+{\rm H.c.}.
\end{equation}
The above Hamiltonian can be rewritten in the following Fourier form
\begin{equation*}
    \hat{H}_{I}(t)=\sum_{n=-\infty}^{\infty}\hat{H}_{n}e^{i n\nu t}, 
\end{equation*}
where
\begin{equation}\label{eq_10}
\small
    \hat{H}_{n}=-\sum_{k}\frac{\Omega_{k} g}{2\Delta_{k}}\mathcal{J}_{n}\left(\frac{\varepsilon}{\nu}\right)e^{-in\phi_{k}}\left[(-1)^{n}\hat{a}\hat{\sigma}_{k}^{sg}+\hat{a}^{\dagger}\hat{\sigma}_{k}^{gs}\right].
\end{equation}
We consider the condition where the external driving frequency is significantly high and far exceeds the characteristic energy inside the system, i.e., $\nu \gg \Omega_{k} g/|\Delta_{k}|$. As a result, introducing the Pauli operators $\hat{\sigma}_{k}^{x}=\hat{\sigma}_{k}^{gs}+\hat{\sigma}_{k}^{sg}$ and $\hat{\sigma}_{k}^{y}\ =i(\hat{\sigma}_{k}^{sg}-\hat{\sigma}_{k}^{gs})$, we can employ perturbative expansion techniques \cite{PhysRevX.4.031027} to derive an approximate effective Hamiltonian by truncating to powers of ($1/\nu$) as follows:
\begin{equation}\label{eq_11}
\small
\begin{split}
    \hat{H}_{\rm eff}=&\hat{H}_{0}+\sum_{n=1}^{\infty}\frac{1}{n\nu}\left[\hat{H}_{n},\hat{H}_{-n}\right] \\
    =&-\sum_{k}\frac{\Omega_{k} g}{2\Delta_{k}}\mathcal{J}_{0}\left(\frac{\varepsilon}{\nu}\right)\left(\hat{a}\hat{\sigma}_{k}^{sg}+\hat{a}^{\dagger}\hat{\sigma}_{k}^{gs}\right) \\
    &+\sum_{k > l}\sum_{n=1}^{\infty}\frac{\Omega_{l}\Omega_{k}g^{2}}{4n\nu\Delta_{l}\Delta_{k}}\mathcal{J}_{n}^{2}\left(\frac{\varepsilon}{\nu}\right)\sin(n\phi_{lk})\left(\hat{\sigma}_{l}^{x}\hat{\sigma}_{k}^{y}-\hat{\sigma}_{l}^{y}\hat{\sigma}_{k}^{x}\right),
\end{split}
\end{equation}
where $\hat{H}_{-n}$ and $\hat{H}_{n}$ satisfy the relation $\hat{H}_{-n}=\hat{H}_{n}^{*}$, and $\phi_{lk}=\phi_{l}-\phi_{k}$. Third-order and even higher-order terms are analyzable using the James’ effective-Hamiltonian method \cite{PhysRevA.95.032124}. For the dispersion region of concern in this study, these terms have minimal impact on the results and can be safely omitted. By adjusting the appropriate ratio of the driving intensity $\varepsilon$ and frequency $\nu$, specifically setting $\varepsilon/\nu = 2.4048$, the first term vanishes and the effective Hamiltonian in Eq. (\ref{eq_11}) can be recast as following form
\begin{equation}\label{H_eff}
    \hat{H}_{\rm eff}=\sum_{k > l}\sum_{n=1}^{\infty}\frac{\Omega_{l}\Omega_{k}g^{2}}{4n\nu\Delta_{l}\Delta_{k}}\mathcal{J}_{n}^{2}\left(\frac{\varepsilon}{\nu}\right)\sin(n\phi_{lk})\left(\hat{\sigma}_{l}^{x}\hat{\sigma}_{k}^{y}-\hat{\sigma}_{l}^{y}\hat{\sigma}_{k}^{x}\right).
\end{equation}
Thus, we derive an effective tunable DMI Hamiltonian with $z$-component. It is evident that when the phase shift equals to zero ($\phi_{lk}=0$), the DMI between site $l$ and site $k$ vanishes. In addition, by adjusting the site-dependent initial phases and the frequencies of the driving fields, we can simulate different DM-type Hamiltonians, and consequently achieve the desired chiral dynamics.


\subsection{Three-state chiral current}
\label{SecIIB}

With controllable parameters to implement the desired dynamic, we now present some examples of three-state chiral current. 

\subsubsection{Three-atom case}
We first begin by examining the scenario where $N=3$ atoms, as this case is crucial for comprehending the chiral dynamics within the system under consideration. We perform the dynamical analysis by choosing proper site-dependent phases with $\phi_{k}=2k\pi/3$ ($k = 1, 2, 3$). The following effective Hamiltonian containing the $z$-component of DMI can be obtained as
\begin{equation}\label{eq_12}
    \hat{H}_{\rm eff}^{(3)}=-\kappa\sum_{k=1}^{3}\left(\hat{\sigma}_{k}^{x}\hat{\sigma}_{k+1}^{y}-\hat{\sigma}_{k}^{y}\hat{\sigma}_{k+1}^{x}\right). 
\end{equation}
Note if $k=3$, then $k+1 \equiv 1$, and the effective coupling strength is
\begin{equation}\label{eq_13}
\small
    \kappa=\frac{\Omega^{2}g^{2}}{4\Delta^{2}\nu}\sum_{n=1}^{\infty}\mathcal{J}_{n}^{2}\left(\frac{\varepsilon}{\nu}\right)\sin\left(\frac{2n\pi}{3}\right)/n, 
\end{equation}
where $\Omega_{k} = \Omega$, $\omega_{k}^{L} = \omega_{L}$ and $\Delta_{k} = \Delta$. Then we can represent the above effective Hamiltonian in its ground-state subspace spanned by \{$\ket{gss}, \ket{sgs}, \ket{ssg}$\} as
\begin{equation}\label{eq_14}
   \hat{H}^{(3)}_{\rm sub} = 2i\kappa\left(\begin{array}{cccc}
0 & -1 & 1 \\
1 & 0 & -1 \\
-1 & 1 & 0
\end{array}\right).
\end{equation} 
The eigenvalues for $\hat{H}^{(3)}_{\rm sub}$ are $\{0, \pm \omega\}$ with $\omega=2\sqrt{3}\kappa$. Therefore, the corresponding time evolution operator can be written as
\begin{equation}\label{eq_15}
   \hat{U}(t) = \left(\begin{array}{cccc}
x_{1}(t) & x_{3}(t) & x_{2}(t) \\
x_{2}(t) & x_{1}(t) & x_{3}(t) \\
x_{3}(t) & x_{2}(t) & x_{1}(t)
\end{array}\right),
\end{equation}
where
\begin{equation}\label{eq_16}
    \begin{split}
        x_{1}(t)&=\frac{1}{3}\left[1+2\cos\left(\omega t\right)\right], \\
        x_{2}(t)&=\frac{1}{3}\left[1+2\cos\left(\omega t-\frac{2}{3}\pi\right)\right], \\
        x_{3}(t)&=\frac{1}{3}\left[1+2\cos\left(\omega t-\frac{4}{3}\pi\right)\right].
    \end{split}
\end{equation}
Initialing the initial state as $\ket{\psi(0)}=\ket{gss}$, and acting the evolution operator $\hat{U}(t)$ on the initial state, we can obtain the time evolution state
\begin{equation}
    \ket{\psi(t)}=x_{1}(t)\ket{gss}+x_{2}(t)\ket{sgs}+x_{3}(t)\ket{ssg}.
\end{equation}
It can be easily verified that 
\begin{equation}
    \begin{split}
        \ket{\psi(t)}= \left\{ {\begin{array}{l}
        {\ket{gss}, \quad  t=2n\pi/(2\sqrt{3}\kappa)} \\
        {\ket{sgs}, \quad t=(2\pi/3+2n\pi)/(2\sqrt{3}\kappa)} \\ 
        {\ket{ssg}, \quad t=(4\pi/3+2n\pi)/(2\sqrt{3}\kappa)}  
\end{array}} \right.
    \end{split}
\end{equation}
Here $n$ is an integer. Thus the state $\ket{g}$ traverses the ring sites in a counterclockwise manner, i.e., $1 \rightarrow 2 \rightarrow 3 \rightarrow 1\cdots$.

Similarly, the time evolution operator in the corresponding ground-state subspace spanned by $\{\ket{sgg}, \ket{gsg}, \ket{ggs}\}$ can be represented as $\hat{U}^{\dagger}(t)$. If we set the initial state as $\ket{\psi(0)}=\ket{sgg}$, the evolution state is 
\begin{equation}
\ket{\psi(t)}=x_{1}(t)\ket{sgg}+x_{3}(t)\ket{gsg}+x_{2}(t)\ket{ggs}.\nonumber
\end{equation}
We can obtain :
\begin{equation}
    \begin{split}
        \ket{\psi(t)}= \left\{ {\begin{array}{l}
        {\ket{sgg}, \quad t=2n\pi/(2\sqrt{3}\kappa)} \\
        {\ket{ggs}, \quad t=(2\pi/3+2n\pi)/(2\sqrt{3}\kappa)} \\ 
        {\ket{gsg}, \quad t=(4\pi/3+2n\pi)/(2\sqrt{3}\kappa)}  
\end{array}} \right.
    \end{split}
\end{equation}
The state $\ket{s}$ traverses the ring sites in a clockwise manner, i.e., $1 \rightarrow 3 \rightarrow 2 \rightarrow 1\cdots$.

For the case more than three atoms, three-state chiral current can be investigated by introducing composite spins. Compared with Ref. \cite{Wang2019}, we can tune the Rabi frequencies and the phases to obtain perfect chiral current. 
\subsubsection{Four-atom case} 

For the case of four atoms, the second and third effective spins can be grouped as a composite spin with $s=1$. Here we choose the initial conditions $\phi_{1} = 2\pi/3, \phi_{2} = \phi_{3} = 4\pi/3, \phi_{4} = 2\pi$ and $\Omega_{1}=\Omega_{4}=\Omega$, $\Omega_{2}=\Omega_{3}=\Omega/\sqrt{2}$. And the corresponding effective Hamiltonian can be written as
\begin{equation}\label{eq_17}
    \hat{H}_{\rm eff}^{(4)}=-\kappa\vec{e}_{z}\cdot\left(\frac{1}{\sqrt{2}}\vec{\sigma}_{1}\times\vec{\Sigma}_{23}+\vec{\sigma}_{4}\times\vec{\sigma}_{1}+\frac{1}{\sqrt{2}}\vec{\Sigma}_{23}\times\vec{\sigma}_{4}\right), 
\end{equation}
where  the composite operator $\vec{\Sigma}_{23} = \vec{\sigma}_{2} + \vec{\sigma}_{3}$ with spin-1. Here we can approximately consider $\Delta_{k}$ to be equal to $\Delta$.
In this situation, we adopt the triplet form to describe the chiral evolution concisely, and the quantum states can be expressed as 
\begin{equation*}
    \begin{split}
        \ket{\mathcal{T}_{-}}&=\ket{ss}, \\
        \ket{\mathcal{T}_{0}}&=\frac{1}{\sqrt{2}}(\ket{gs}+\ket{sg}), \\
        \ket{\mathcal{T}_{+}}&=\ket{gg}.
    \end{split}
\end{equation*}
The representation of the Hamiltonian in Eq. (\ref{eq_17}) in the ground-state subspace spanned by $\{\ket{g \mathcal{T}_{-} s}, \ket{s \mathcal{T}_{0} s}, \ket{s \mathcal{T}_{-} g}\}$ is identical to that in Eq. (\ref{eq_14}). Similarly, within the ground-state subspace spanned by $\{\ket{s \mathcal{T}_{+} g}, \ket{g \mathcal{T}_{0} g}, \ket{g \mathcal{T}_{+} s}\}$, the time evolution operator can also be denoted as $\hat{U}^{\dagger}(t)$.
\subsubsection{Five-atom case } 

For the case of five atoms, we set three different modulation phases. The first type of phase distribution can modulate the interaction between a site with spin-1/2 and two composite sites with spin-1. The phases are set to be $\phi_{1}=2\pi/3, \phi_{2}=\phi_{3}=4\pi/3, \phi_{4}=\phi_{5}=2\pi$ and $\Omega_{1}=\Omega, \Omega_{k}=\Omega/\sqrt{2}$ with $k=2, 3, 4, 5$. And the system evolution is governed by the effective Hamiltonian as follows:
\begin{equation}\label{eq_18}
    \hat{H}_{\rm eff}^{(5,1)}=-\kappa\vec{e}_{z}\cdot\left(\frac{1}{\sqrt{2}}\vec{\sigma}_{1}\times\vec{\Sigma}_{23}+\frac{1}{\sqrt{2}}\vec{\Sigma}_{45}\times\vec{\sigma}_{1}+\frac{1}{2}\vec{\Sigma}_{23}\times\vec{\Sigma}_{45}\right), 
\end{equation}
where $\vec{\Sigma}_{45} = \vec{\sigma}_{4} + \vec{\sigma}_{5}$. 
The representation of the Hamiltonian in Eq. (\ref{eq_18}) within the ground-state subspace spanned by $\{\ket{g \mathcal{T}_{-} \mathcal{T}_{-}}, \ket{s \mathcal{T}_{0} \mathcal{T}_{-}}, \ket{s\mathcal{T}_{-} \mathcal{T}_{0}}\}$ coincides with that presented in Eq. (\ref{eq_14}). Likewise, within the ground-state subspace spanned by $\{\ket{s \mathcal{T}_{+} \mathcal{T}_{+}}, \ket{g \mathcal{T}_{0} \mathcal{T}_{+}}, \ket{g\mathcal{T}_{+} \mathcal{T}_{0}}\}$, the time evolution operator can be equivalently recast as $\hat{U}^{\dagger}(t)$.

Similarly, the second type can be regarded as the interaction of spin-3/2 with two spin-1/2 particles. If we initialize the phase to $\phi_{1}=2\pi/3, \phi_{2}=4\pi/3, \phi_{3}=\phi_{4}=\phi_{5}=2\pi$ and $\Omega_{1}=\Omega_{2}=\Omega, \Omega_{3}=\Omega_{4}=\Omega_{5}=\Omega/\sqrt{3}$, the effective Hamiltonian can be written as
\begin{equation}\label{eq_19}
    \hat{H}_{\rm eff}^{(5,2)}=-\kappa\vec{e}_{z}\cdot\left(\vec{\sigma}_{1}\times\vec{\sigma}_{2}+\frac{1}{\sqrt{3}}\vec{\sigma}_{2}\times\vec{\Sigma}_{345}+\frac{1}{\sqrt{3}}\vec{\Sigma}_{345}\times\vec{\sigma}_{1}\right), 
\end{equation}
where $\vec{\Sigma}_{345} = \vec{\sigma}_{3} +\vec{\sigma}_{4} + \vec{\sigma}_{5}$. For this situation, we represent the quantum state using a quartet notation, where
\begin{equation*}
    \begin{split}
       \ket{\mathcal{Q}_{-\frac{3}{2}}}&=\ket{sss}, \\
       \ket{\mathcal{Q}_{-\frac{1}{2}}}&=\frac{1}{\sqrt{3}}(\ket{gss}+\ket{sgs}+\ket{ssg}), \\
       \ket{\mathcal{Q}_{+\frac{1}{2}}}&=\frac{1}{\sqrt{3}}(\ket{sgg}+\ket{gsg}+\ket{ggs}), \\
       \ket{\mathcal{Q}_{+\frac{3}{2}}}&=\ket{ggg}.
    \end{split}
\end{equation*}
Similar to the previous discussion, the form of the Hamiltonian in Eq. (\ref{eq_19}) within the ground-state subspace spanned by $\{\ket{gs\mathcal{Q}_{-\frac{3}{2}}}, \ket{sg\mathcal{Q}_{-\frac{3}{2}}}, \ket{ss\mathcal{Q}_{-\frac{1}{2}}}\}$ is given by Eq. (\ref{eq_14}). Additionally, the time evolution operator in the other subspace spanned by $\{\ket{sg\mathcal{Q}_{\frac{3}{2}}}, \ket{gs\mathcal{Q}_{\frac{3}{2}}}, \ket{gg\mathcal{Q}_{\frac{1}{2}}}\}$ can be represented as $\hat{U}^{\dagger}$. In summary, the evolution direction of the three-state chiral currents are shown in Table \ref{tab-1}. 
\begin{table}[h]
\centering
\caption{The evolution direction of the quantum states versus time.}
\label{tab-1}
\setlength{\tabcolsep}{1mm}
\begin{tabular}{c|cccc}
\hline
\hline
 \diagbox [width=4em,trim=l]{$N$}{$t$} & 0 & $\pi/(3\sqrt{3}\kappa)$ & $2\pi/(3\sqrt{3}\kappa)$ & $\pi/(\sqrt{3}\kappa)$ \\
\hline
   \multirow{2}{*}{$3$}& 
   $\ket{gss}$ & $\ket{sgs}$ & $\ket{ssg}$ & 
   $\ket{gss}$ \\
   & $\ket{sgg}$ & $\ket{ggs}$ & $\ket{gsg}$ & $\ket{sgg}$ \\
   \hline
  \multirow{2}{*}{$4$} & $\ket{g\mathcal{T}_{-}s}$ & $\ket{s\mathcal{T}_{0}s}$ & $\ket{s\mathcal{T}_{-}g}$ & $\ket{g\mathcal{T}_{-}s}$ \\
   & $\ket{s\mathcal{T}_{+}g}$ & $\ket{g\mathcal{T}_{+}s}$ & $\ket{g\mathcal{T}_{0}g}$ & $\ket{s\mathcal{T}_{+}g}$ \\
   \hline
   \multirow{4}{*}{$5$}& $\ket{g\mathcal{T}_{-}\mathcal{T}_{-}}$ & $\ket{s\mathcal{T}_{0}\mathcal{T}_{-}}$ & $\ket{s\mathcal{T}_{-}\mathcal{T}_{0}}$ & $\ket{g\mathcal{T}_{-}\mathcal{T}_{-}}$ \\
   & $\ket{s\mathcal{T}_{+}\mathcal{T}_{+}}$ & $\ket{g\mathcal{T}_{+}\mathcal{T}_{0}}$ & $\ket{g\mathcal{T}_{0}\mathcal{T}_{+}}$ & $\ket{s\mathcal{T}_{+}\mathcal{T}_{+}}$ \\
   & $\ket{gs\mathcal{Q}_{-\frac{3}{2}}}$ & $\ket{sg\mathcal{Q}_{-\frac{3}{2}}}$ & $\ket{ss\mathcal{Q}_{-\frac{1}{2}}}$ & $\ket{gs\mathcal{Q}_{-\frac{3}{2}}}$ \\
   & $\ket{sg\mathcal{Q}_{\frac{3}{2}}}$ & $\ket{gg\mathcal{Q}_{\frac{1}{2}}}$ & $\ket{gs\mathcal{Q}_{\frac{3}{2}}}$ & $\ket{sg\mathcal{Q}_{\frac{3}{2}}}$ \\
\hline
\hline
\end{tabular}
\end{table}

\subsection{Multi-state chiral current}
\label{SecIIC}

In addition to the aforementioned dynamics evolution, we also modulate the third type of phase distribution with $\phi_{k}=-2k\pi/5$.
The corresponding effective Hamiltonian reads
\begin{equation}\label{eq_20}
    \hat{H}_{\rm eff}^{(5,3)}=\kappa_{1}\hat{o}_{1}+\kappa_{2}\hat{o}_{2}, 
\end{equation}
where
\begin{subequations}\label{eq_21}
    \begin{align}
    \hat{o}_{1}&=\sum_{k=1}^{5}\vec{e}_{z}\cdot\left(\vec{\sigma}_{k}\times\vec{\sigma}_{k+1}\right), \label{eq_21c} \\
      \hat{o}_{2}&=\sum_{k=1}^{5}\vec{e}_{z}\cdot\left(\vec{\sigma}_{k}\times\vec{\sigma}_{k+2}\right). \label{eq_21d} 
    \end{align}
\end{subequations}
Here we employ the periodic boundary condition, which implies $k \equiv$ ($k$ mod $N$). The Hamiltonian in Eq. (\ref{eq_20}) contains both nearest-neighbor and next-nearest-neighbor terms.
Indeed, it is necessary to satisfy the condition  $\kappa_{2}/\kappa_{1}=\sqrt{5}-2 \approx 0.2361$ \cite{Wang2019}, we can obtain the perfect ground-state chiral current in this situation. Subsequently, we will proceed with a brief analysis of its dynamic process.

In the ground-state subspace spanned by $\{\ket{gssss}, \ket{sgsss}, \ket{ssgss}, \ket{sssgs}, \ket{ssssg}\}$, the Hamiltonian in Eq. (\ref{eq_20}) is represented as
\begin{equation}
    \hat{H}_{\rm sub}^{(5,3)} = \frac{i}{2}\left( {\begin{array}{*{20}{c}}
{0}&{\kappa_{1}}&{\kappa_{2}}&{-\kappa_{2}}&{-\kappa_{1}}\\
{-\kappa_{1}}&{0}&{\kappa_{1}}&{\kappa_{2}}&{-\kappa_{2}}\\
{-\kappa_{2}}&{-\kappa_{1}}&{0}&{\kappa_{1}}&{\kappa_{2}}\\
{\kappa_{2}}&{-\kappa_{2}}&{-\kappa_{1}}&{0}&{\kappa_{1}}\\
{\kappa_{1}}&{\kappa_{2}}&{-\kappa_{2}}&{-\kappa_{1}}&{0}
\end{array}} \right).
\end{equation}
The eigenvalues for $\hat{H}_{\rm eff}^{(5,3)}$ are $\{0, \pm \omega, \pm 3\omega\}$ with $\omega=\sqrt{5-2\sqrt{5}}\kappa_{1}/2 \approx 0.3633\kappa_{1}$. Accordingly, the time evolution operator can be recast as
\begin{equation}\label{eq_23}
   \hat{U}'(t) = \left(\begin{array}{ccccc}
y_{1}(t) &
y_{5}(t) & y_{4}(t) & y_{3}(t) & y_{2}(t) \\
y_{2}(t) & y_{1}(t) & y_{5}(t) & y_{4}(t) & y_{3}(t) \\
y_{3}(t) & y_{2}(t) & y_{1}(t) & y_{5}(t) & y_{4}(t) \\
y_{4}(t) & y_{3}(t) & y_{2}(t) & y_{1}(t) & y_{5}(t) \\
y_{5}(t) & y_{4}(t) & y_{3}(t) & y_{2}(t) & y_{1}(t)
\end{array}\right),
\end{equation}
where
\begin{equation}\label{eq_24}
    \begin{split}
        y_{1}(t)&=\frac{1}{5}\left[1+4\cos(2\omega t)\cos(\omega t)\right], \\
        y_{2}(t)&=\frac{1}{5}\left[1+4\cos\left(2\omega t+\frac{3}{5}\pi\right)\cos\left(\omega t-\frac{1}{5}\pi\right)\right], \\
        y_{3}(t)&=\frac{1}{5}\left[1+4\cos\left(2\omega t+\frac{1}{5}\pi\right)\cos\left(\omega t+\frac{3}{5}\pi\right)\right], \\
        y_{4}(t)&=\frac{1}{5}\left[1+4\cos\left(2\omega t+\frac{4}{5}\pi\right)\cos\left(\omega t+\frac{2}{5}\pi\right)\right], \\
        y_{5}(t)&=\frac{1}{5}\left[1+4\cos\left(2\omega t+\frac{7}{5}\pi\right)\cos\left(\omega t+\frac{1}{5}\pi\right)\right].
    \end{split}
\end{equation}
When the initial state is $\ket{gssss}$, the evolution state reads
\begin{equation}
\begin{split}
    \ket{\psi(t)}=&y_{1}(t)\ket{gssss}+y_{2}(t)\ket{sgsss}+y_{3}(t)\ket{ssgss} \\
    &+y_{4}(t)\ket{sssgs}+y_{5}(t)\ket{ssssg}.
\end{split}
\end{equation}
When the evolution time $t = 0$, $2\pi/(5\omega)$, $4\pi/(5\omega)$, $6\pi/(5\omega)$, $8\pi/(5\omega)$, $2\pi/(\omega)$, the evolution state $\ket{\psi(t)}=\ket{gssss}$, $\ket{ssgss}$, $\ket{ssssg}$, $\ket{sgsss}$, $\ket{sssgs}$, $\ket{gssss}$, respectively.
The ground state $\ket{g}$ traverses the ring sites in the direction of $1 \rightarrow 3 \rightarrow 5 \rightarrow 2 \rightarrow 4 \rightarrow 1$ within one period of $T=2\pi/\omega$.
In the ground-state subspace spanned by $\{\ket{sgggg}, \ket{gsggg}, \ket{ggsgg}, \ket{gggsg}, \ket{ggggs}\}$, the time evolution operator, denoted as $\hat{U}'^{\dagger}(t)$, gives rise to an opposite evolution direction, i.e., $1 \rightarrow 4 \rightarrow 2 \rightarrow 5 \rightarrow 3 \rightarrow 1\cdots$.

For the case $\varepsilon/\nu=2.4048$, the effective coupling constants in the Hamiltonian in Eq. (\ref{eq_20}) can be reduced to
\begin{equation}
    \begin{split}
      \kappa_{1}&=\frac{\Omega^{2}g^{2}}{4\Delta^{2}\nu}\sum_{n=1}^{\infty}\mathcal{J}_{n}^{2}\left(\frac{\varepsilon}{\nu}\right)\sin\left(\frac{2n\pi}{5}\right)/n, \\
      \kappa_{2}&=\frac{\Omega^{2}g^{2}}{4\Delta^{2}\nu}\sum_{n=1}^{\infty}\mathcal{J}_{n}^{2}\left(\frac{\varepsilon}{\nu}\right)\sin\left(\frac{4n\pi}{5}\right)/n.
    \end{split}
\end{equation}
with $\Omega_{k} = \Omega$ and $\Delta_{k} = \Delta$. Unfortunately, the ratio of coupling constant $\kappa_{2}/\kappa_{1}$ for such specific case is $0.2702$, so we can obtain the imperfect ground state chirality, which will be illustrated numerically in \ref{SecIIIB}.



\subsection{Simulating DMI by Trotter-Suzuki formulas}
\label{SecIID}
Recently, several studies have utilized Floquet Hamiltonian engineering to design periodic drive sequences for obtaining the required effective Hamiltonian in the system \cite{PhysRevX.10.031002,doi:10.1126/science.abd9547,PRXQuantum.3.020303,PRXQuantum.4.010334,PhysRevApplied.18.024033}. This approach can enable us to perform operations that are unattainable through conventional methods, thereby enhancing the efficiency and reliability of quantum computing and quantum information processing. Already at this point, we have successfully obtained the $z$-component of DMI. Therefore, we can design the pulse sequence by Floquet Hamiltonian engineering and employ the first-order or more accurate second-order Trotter-Suzuki expansion \cite{2020PhRvA.101e2333Z,gong2023improved} to simulate the complete DMI as much as possible. 

Based on the first-order Trotter-Suzuki expansion
\begin{equation}    \hat{\mathcal{U}}_{1}=\prod\limits_{j=n}^{1} {\exp\left(-i\hat{\mathcal{H}}'_{j}\tau_{j}\right)},
\end{equation}
the time evolution operator for one period under the toggling frame can be written as
\begin{equation}\label{B1}
    \begin{split}
        \hat{\mathcal{U}}_{1}&=\hat{\mathcal{R}}_{n}e^{-i\hat{\mathcal{H}}_{n}\tau_{n}}\cdots\hat{\mathcal{R}}_{1}e^{-i\hat{\mathcal{H}}_{1}\tau_{1}} \\
        &=\left(\hat{\mathcal{R}}_{n}\cdots\hat{\mathcal{R}}_{1}\right)e^{-i\hat{\mathcal{H}}'_{n}\tau_{n}}\cdots e^{-i\hat{\mathcal{H}}'_{1}\tau_{1}} \\
        &=e^{-i\hat{\mathcal{H}}'_{n}\tau_{n}}\cdots e^{-i\hat{\mathcal{H}}'_{1}\tau_{1}}, 
    \end{split}
\end{equation}
where
\begin{equation}\label{B2}
    \hat{\mathcal {H}}'_{n}=\left(\hat{\mathcal{R}}_{n-1}\cdots\hat{\mathcal {R}}_{1}\right)^{-1}\hat{\mathcal {H}}_{n}\left(\hat{\mathcal {R}}_{n-1}\cdots\hat{\mathcal {R}}_{1}\right).
\end{equation}
Since the operator acts in the rotated frame determined by the previous pulses, we usually call it as “toggling frame” \cite{PRXQuantum.4.010334}. And the condition $\hat{\mathcal{R}}_{n}\cdots\hat{\mathcal{R}}_{1}=\mathcal{I}$ needs to be satisfied while designing the pulse sequence, which ensures that the frame can rotate back to the original frame at the end of a period. 
\begin{figure}[htbp]
    \centering
\includegraphics[angle=0,width=0.48\textwidth]{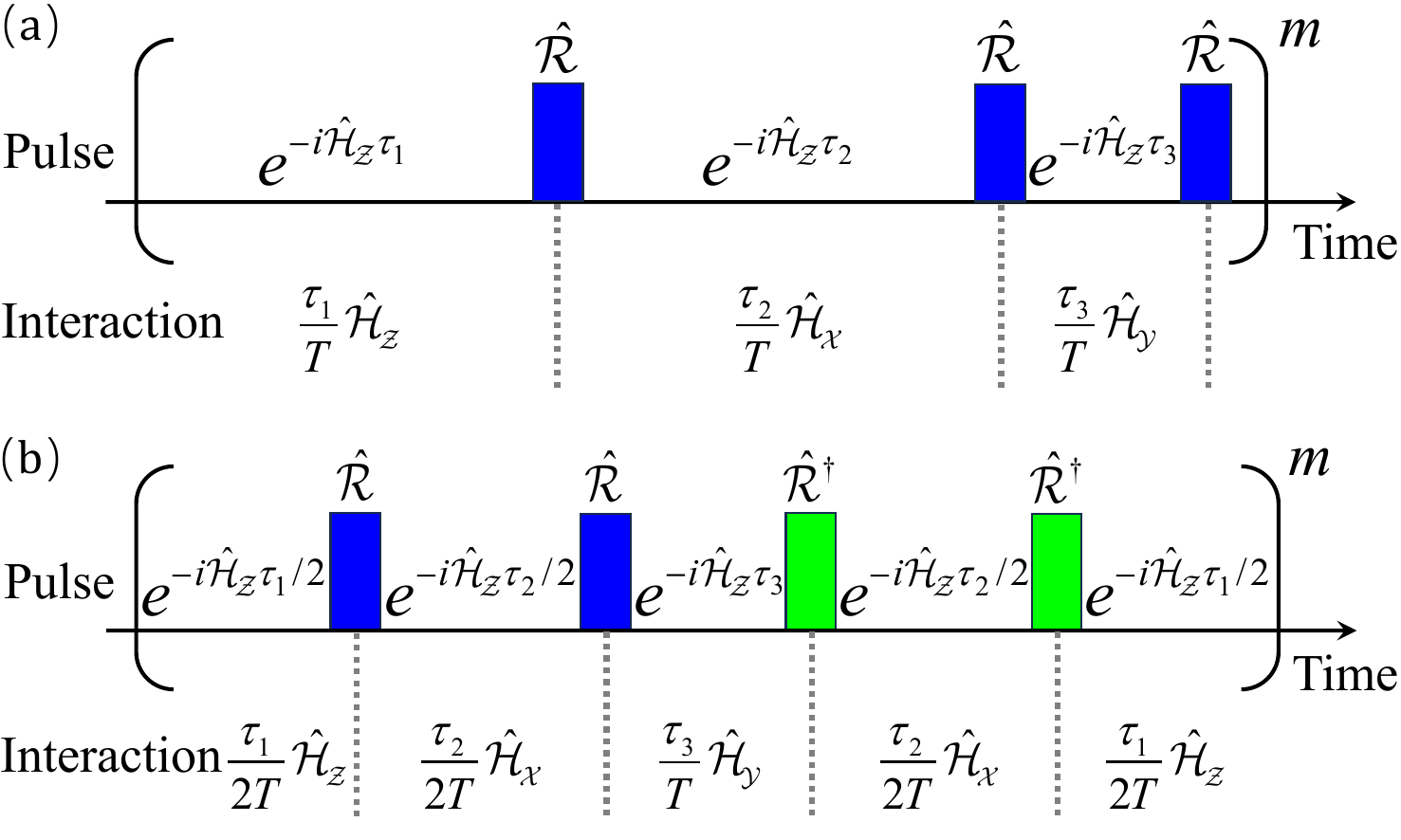}
    \caption{(a) The pulse sequence ($\hat{\mathcal{R}}=\hat{\mathcal{P}}_{1}\hat{\mathcal{P}}_{2}\hat{\mathcal{P}}_{3}$) of DMI by using the first-order Trotter-Suzuki expansion with $T=\tau_{1}+\tau_{2}+\tau_{3}$. $\hat{\mathcal{H}}_{\mathcal{X}}$, $\hat{\mathcal{H}}_{\mathcal{Y}}$, $\hat{\mathcal{H}}_{\mathcal{Z}}$ represent the three components of DMI. (b) The pulse sequence of DMI based on the second-order Trotter-Suzuki expansion.}
    \label{fig02}
\end{figure}

The unitary evolution operator can also be interpreted as an approximation result of the average Hamiltonian theory \cite{https://doi.org/10.1002/cmr.a.21414}
\begin{equation}\label{B3}
    \hat{\mathcal{U}}_{1}\approx e^{-i\bar{\mathcal{H}}T}, 
\end{equation}
with
\begin{equation}\label{B4}
\bar{\mathcal{H}}=\frac{1}{T}\sum_{j=1}^{n}\hat{\mathcal{H}}'_{j}\tau_{j},
\end{equation}
where $\tau_{j} (j=1,2,\cdots n)$ are the free evolution time of the system between pulse applications and we ignore the pulse time here. $T$ is the evolution time of one period, and we approximate the target evolution by continuously repeating the operations within one period. After $m$ cycles of such period, the time evolution operator can be expressed as $\hat{\mathcal{U}}=(\hat{\mathcal{U}}_{1})^{m}$. When $m$ is larger, the obtained Hamiltonian will be closer to the target Hamiltonian. 

The protocol based on the first-order Trotter-Suzuki expansion can be improved by considering the second-order Trotter-Suzuki expansion
\begin{equation}
    \hat{\mathcal{U}}_{2}=\prod\limits_{j = 1}^{n} {\exp\left(-i\hat{\mathcal{H}}'_{j}\frac{\tau_{j}}{2}\right)}\prod\limits_{j = n}^{1} {\exp\left(-i\hat{\mathcal{H}}'_{j}\frac{\tau_{j}}{2}\right)}.
\end{equation}
The time evolution operator for one cycle can be rewritten as
\begin{widetext}
\begin{equation}
    \begin{split}
        \hat{\mathcal{U}}_{2}=&e^{-i\hat{\mathcal{H}}_{1}\frac{\tau_{1}}{2}}\hat{\mathcal{R}}^{\dagger}_{1} \cdots e^{-i\hat{\mathcal{H}}_{n}\frac{\tau_{n}}{2}}\hat{\mathcal{R}}_{n}^{\dagger} \cdot
        \hat{\mathcal{R}}_{n}e^{-i\hat{\mathcal{H}}_{n}\frac{\tau_{n}}{2}}\cdots\hat{\mathcal{R}}_{1}e^{-i\hat{\mathcal{H}}_{1}\frac{\tau_{1}}{2}} \\
        =&e^{-i\hat{\mathcal{H}}'_{1}\frac{\tau_{1}}{2}} \cdots e^{-i\hat{\mathcal{H}}'_{n}\frac{\tau_{n}}{2}}\left(\hat{\mathcal{R}}_{1}^{\dagger}\cdots\hat{\mathcal{R}}_{n}^{\dagger}\right) \cdot \left(\hat{\mathcal{R}}_{n}\cdots\hat{\mathcal{R}}_{1}\right)e^{-i\hat{\mathcal{H}}'_{n}\frac{\tau_{n}}{2}}\cdots e^{-i\hat{\mathcal{H}}'_{1}\frac{\tau_{1}}{2}} \\
        =&e^{-i\hat{\mathcal{H}}'_{1}\frac{\tau_{1}}{2}} \cdots e^{-i\hat{\mathcal{H}}'_{n}\frac{\tau_{n}}{2}} \cdot e^{-i\hat{\mathcal{H}}'_{n}\frac{\tau_{n}}{2}}\cdots e^{-i\hat{\mathcal{H}}'_{1}\frac{\tau_{1}}{2}} \\
        \approx& e^{-i\bar{\mathcal{H}}T}.
    \end{split}
\end{equation}
\end{widetext}
Then we can obtain the target Hamiltonian in Eq. (\ref{B4}) more accurately.

Based on the fundamental theory mentioned above, we can simulate the complete DMI using the obtained $z$-component in the Hamiltonian (\ref{eq_12}) that we call $\hat{\mathcal{H}}_{\mathcal{Z}}$ here. With the properties of Pauli operators, we can introduce the transformation $\hat{\mathcal{P}}_{k}=\exp(i\frac{\pi}{4}\hat{\sigma}^{y}_{k})\exp(i\frac{\pi}{4}\hat{\sigma}^{x}_{k}) (k=1, 2, 3)$ and further derive the following relationship
\begin{equation}   \hat{\mathcal{P}}^{\dagger}_{k}\hat{\sigma}^{x}_{k}\hat{\mathcal{P}}_{k}=\hat{\sigma}^{y}_{k}, \ \ \ \hat{\mathcal{P}}^{\dagger}_{k}\hat{\sigma}^{y}_{k}\hat{\mathcal{P}}_{k}=\hat{\sigma}^{z}_{k}, \ \ \ \hat{\mathcal{P}}^{\dagger}_{k}\hat{\sigma}^{z}_{k}\hat{\mathcal{P}}_{k}=\hat{\sigma}^{x}_{k}.
\end{equation}
Then we can obtain the other two components of DMI
\begin{equation}
    \begin{split}
       \hat{\mathcal{H}}_{\mathcal{X}}=\hat{\mathcal{R}}^{\dagger}\hat{\mathcal{H}}_{\mathcal{Z}}\hat{\mathcal{R}}=-\kappa\sum_{k=1}^{3}\left(\hat{\sigma}_{k}^{y}\hat{\sigma}_{k+1}^{z}-\hat{\sigma}_{k}^{z}\hat{\sigma}_{k+1}^{y}\right), \\
       \hat{\mathcal{H}}_{\mathcal{Y}}=\hat{\mathcal{R}}^{\dagger}\hat{\mathcal{H}}_{\mathcal{X}}\hat{\mathcal{R}}=-\kappa\sum_{k=1}^{3}\left(\hat{\sigma}_{k}^{z}\hat{\sigma}_{k+1}^{x}-\hat{\sigma}_{k}^{x}\hat{\sigma}_{k+1}^{z}\right),
    \end{split}
\end{equation}
where $\hat{\mathcal{R}}=\hat{\mathcal{P}}_{1}\hat{\mathcal{P}}_{2}\hat{\mathcal{P}}_{3}$. Ultimately, based on the first-order Trotter-Suzuki expansion, 
we can obtain the following target Hamiltonian by successively applying three pulses $\hat{\mathcal{R}}$ to the sequence as shown in Fig. \ref{fig02}(a), which represents the weighted average of the Hamiltonians for each toggling frame
\begin{equation}\label{B5}
    \bar{\mathcal{H}}=\frac{1}{T}\left(\tau_{2}\hat{\mathcal{H}}_{\mathcal{X}}+\tau_{3}\hat{\mathcal{H}}_{\mathcal{Y}}+\tau_{1}\hat{\mathcal{H}}_{\mathcal{Z}}\right).
\end{equation}
Here $\tau_{1}+\tau_{2}+\tau_{3}=T$, and the symbols $\hat{\mathcal{H}}_{\mathcal{X}}$, $\hat{\mathcal{H}}_{\mathcal{Y}}$, and $\hat{\mathcal{H}}_{\mathcal{Z}}$ denote three components of DMI in the $x$, $y$, $z$ directions respectively. The time duration $\tau_{n}$ control the coupling strength of the corresponding interaction terms. For our case, the coupling strength of $x$, $y$, $z$ components are $\tau_{2}\kappa/T$, $\tau_{3}\kappa/T$, $\tau_{1}\kappa/T$, respectively.

Following the second-order Trotter-Suzuki expansion, we can derive the Hamiltonian in Eq. (\ref{B5}) precisely by sequentially applying two pulses $\hat{\mathcal{R}}$ and $\hat{\mathcal{R}}^{\dagger}$ as illustrated in Fig. \ref{fig02}(b). Specifically, the second-order Trotter-Suzuki formula reduces errors by decomposing the evolution operator into more refined steps. With each time step, it visits each Hamiltonian multiple times under the toggling frame, whereas the first-order Trotter-Suzuki formula visit only once. Such complex process allows to approximate the evolution of a quantum system more effectively, thereby enhancing the accuracy and stability of simulations \cite{PhysRevLett.128.210501}.

\section{Numerical simulation of ground-state chiral current}
\label{SecIII}
In this section, we provide a detailed analysis of dissipative effects by considering the influence of the spontaneous emission of atoms and present experimentally feasible parameters for implementing our proposed scheme. Subsequently, the numerical simulation results continue to show that ground-state chiral current can be completely achieved.

\subsection{Decoherence and experimental feasibility}
\label{SecIIIA}
\begin{figure}
    \centering
    \includegraphics[width=0.48\textwidth]{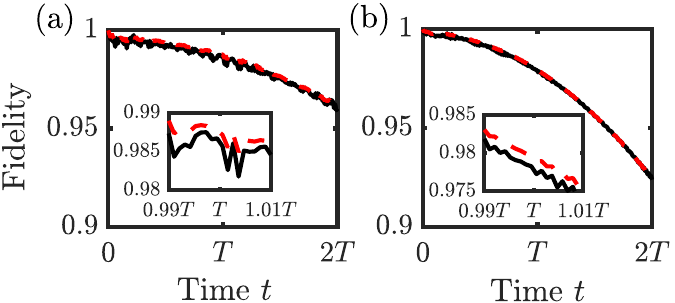}
    \caption{(a) and (b) show the fidelity of the chiral current generated from $\ket{gss}$ and $\ket{sgg}$ versus time, including the red dashed line $\mathcal{F} = |\langle \tilde{\psi}(t)| \psi(t) \rangle|^{2}$ and black line $\mathcal{F} = |\langle \psi(t)|\hat{\rho}|\psi(t) \rangle|$, where $\ket{\tilde{\psi}(t)}$, $\hat{\rho}$ and $\ket{\psi(t)}$ are the evolution states of the full Hamiltonian in Eq. (\ref{eq_03}) without considering dissipative processes, the full Hamiltonian in Eq. (\ref{eq_26}) under dissipative processes with $\gamma=2\pi \times 0.005$ MHz and the effective Hamiltonian in Eq. (\ref{eq_12}), respectively. And $T = \pi/\sqrt{3}\kappa \approx 41.79 \mu s$.}
    \label{fig03}
\end{figure}

To simplify the analysis, we consider the effects of atomic spontaneous emission by assuming that the excited state $\ket{e}$ decays into the ground states $\ket{g}$ and $\ket{s}$ with the same spontaneous emission rate. Therefore, taking the case of three atoms as an example, the dynamics of the system can be described by the Markovian master equation
\begin{equation}\label{eq_26}
    \frac{d\hat{\rho}}{dt}=-i[\hat{\tilde{H}}(t),\hat{\rho}]+\sum_{k=1}^{6}\hat{\mathcal{L}_{k}}\hat{\rho}\hat{\mathcal{L}}_{k}^{\dagger}-\frac{1}{2}\{\hat{\mathcal{L}}_{k}^{\dagger}\hat{\mathcal{L}}_{k},\hat{\rho}\},
\end{equation}
where the dissipative operators are given as follows
\begin{equation*}
 \begin{split}
   &\hat{\mathcal{L}}_{1}=\sqrt{\gamma/2}\ket{g_{1}}\bra{e_{1}},\quad  \hat{\mathcal{L}}_{2}=\sqrt{\gamma/2}\ket{s_{1}}\bra{e_{1}}, \\
   &\hat{\mathcal{L}}_{3}=\sqrt{\gamma/2}\ket{g_{2}}\bra{e_{2}},\quad  \hat{\mathcal{L}}_{4}=\sqrt{\gamma/2}\ket{s_{2}}\bra{e_{2}}, \\
   &\hat{\mathcal{L}}_{5}=\sqrt{\gamma/2}\ket{g_{3}}\bra{e_{3}},\quad  \hat{\mathcal{L}}_{6}=\sqrt{\gamma/2}\ket{s_{3}}\bra{e_{3}}.
 \end{split}
\end{equation*}
The notation $\{\hat{A},\hat{B}\} \equiv \hat{A}\hat{B}+\hat{B}\hat{A}$ is the anticommutator of operators $\hat{A}$ and $\hat{B}$. In Fig. \ref{fig03}, we can confirm the accuracy of the Hamiltonian in Eq. (\ref{eq_03}), indicating that the spontaneous atomic emission has a minimal impact on the temporal evolution of the system. Thus, in the following text, we can safely disregard the influence of decoherence.

We investigate the spin dynamics of trapped Rb-87 atoms in a single-mode optical resonator. The conduit for mediating interactions from the $\ket{5S_{1/2}} \rightarrow \ket{5P_{3/2}}$ transitions is a 780.02-nm cavity mode at detuning as high as several tens of GHz \cite{PhysRevLett.122.010405}. The energy level splitting of the hyperfine structure between levels $\ket{5S_{1/2}, F=2}$ and $\ket{5S_{1/2}, F=1}$ is approximately 6834.68 MHz. The carrier frequency $\omega_{L}$ roughly equals to 384 THz \cite{PhysRevResearch.4.013089}. In this work, we choose the parameter $\omega_{g}=2\pi \times 1087.773$ MHz, $\omega_{e}=2\pi \times 61.171$ THz, $\omega_{L}=2\pi \times 61.163$ THz, $\Delta=2\pi \times 8.1$ GHz, $\Omega=2\pi \times 405$ MHz, $g=2\pi \times 90$ MHz and $\nu=2\pi \times 112.5$ MHz uniformly. Thus, all the necessary conditions  for achieving effective dynamics have been successfully satisfied.
\begin{figure}[htbp]
    \centering
    \includegraphics[width=0.48\textwidth]{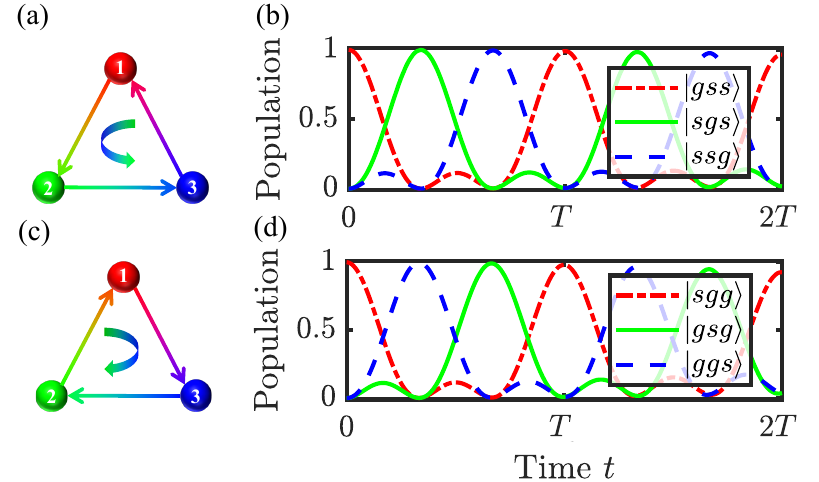}
    \caption{The schematic and population diagrams illustrate the evolution of chirality versus time generated from the states $\ket{gss}$ and $\ket{sgg}$ along the anticlockwise and the clockwise directions, respectively. The dynamics is governed by the Hamiltonian in Eq. (\ref{eq_06}).}
    \label{fig04}
\end{figure}

\subsection{Numerical simulation of three-state and multi-state chirality}
\label{SecIIIB}

Floquet engineering is a typical strategy for controlling long-term system dynamics, characterized by periodically modulating the system frequency to synthesize the target Hamiltonian \cite{PhysRevX.4.031027}. Theoretical studies have demonstrated that Floquet modulation can be employed to generate arbitrary two-body spin interactions  \cite{PhysRevApplied.9.064029,PhysRevA.99.012333}. By periodically adjusting the system parameters, we have successfully synthesized the $z$-component of DMI. Moving forward, we conduct numerical simulations to investigate the chiral dynamics involving three, four, and five atoms.

In what follows, we first present an example of three atoms arranged in a triangular loop to discuss the chiral current with initial state $\ket{gss}$. We provide numerical simulations based on the Hamiltonian (\ref{eq_06}) and present the chiral dynamics depicted in Figs. \ref{fig04}(a) and \ref{fig04}(b). The results reveal an interesting dynamical pattern: the state $\ket{g}$ injected into one of the sites flows anticlockwise through the three-atom sites within one period of $T = \pi/\sqrt{3}\kappa \approx 41.79 \mu s$. Such phenomenon totally aligns with previous theoretical prediction. Furthermore, when considering $\ket{sgg}$ as the initial state, the transfer of the state $\ket{s}$ follows the clockwise rotation $1 \rightarrow 3 \rightarrow 2 \rightarrow 1 \cdots$ in Figs. \ref{fig04}(c) and \ref{fig04}(d). More remarkably, the initial state flip causes the opposite direction of chiral evolution, which can be attributed to the preservation of time reversal symmetry by the $z$-component of DMI.
\begin{figure}
    \centering
    \includegraphics[width=0.48\textwidth]{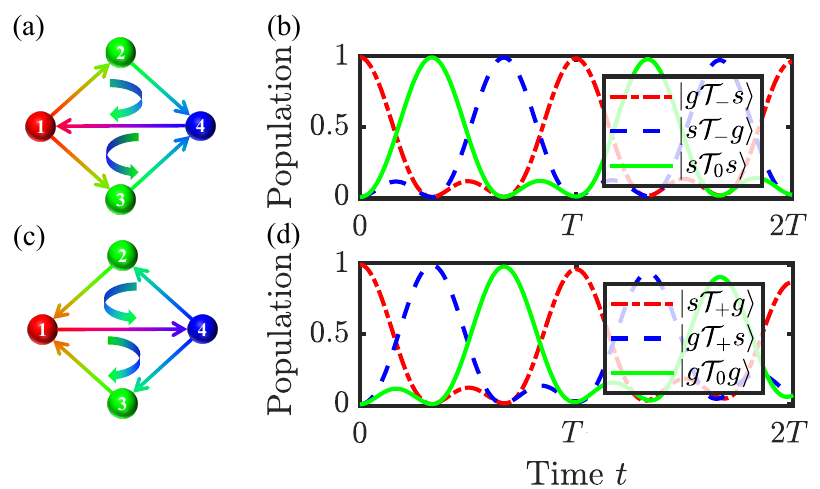}
    \caption{The schematic and population diagrams illustrate the evolution of chirality versus time generated from the states  $\ket{g\mathcal{T}_{-}s}$ and $\ket{s\mathcal{T}_{+}g}$, respectively. The dynamics is governed by the Hamiltonian in Eq. (\ref{eq_06}).}
    \label{fig05}
\end{figure}

The abundance of phases affords us the opportunity to witness a diverse range of chiral dynamics across multiple atoms. In Fig. \ref{fig05}, we illustrate the population transfer of four atoms. Obviously, the numerical results for the initial states $\ket{g\mathcal{T}_{-}s}$ and $\ket{s\mathcal{T}_{+}g}$ exhibit the opposite chirality. In other words, when the initial state is $\ket{g\mathcal{T}_{-}s}$, the evolutionary path of the state $\ket{g}$ follows a sequence of $1 \rightarrow \{2, 3\} \rightarrow 4 \rightarrow 1\cdots$. Otherwise, the state $\ket{s}$ circulates along the route $1 \rightarrow 4 \rightarrow \{2, 3\} \rightarrow 1\cdots$ in the scenario of flipping the initial state.
\begin{table}[b]
\centering
\caption{The fidelity of chiral current generated from different ground states for different $N$-atoms versus time $t$, such as $\mathcal{F}_{\ket{gss}} = |\langle \tilde{\psi}(t)| \psi(t) \rangle|^{2}$, where $\ket{\tilde{\psi}(t)}$ and $\ket{\psi(t)}$ are the evolution states of the full Hamiltonian in Eq. (\ref{eq_06}) and the effective Hamiltonian in Eq. (\ref{eq_12}), respectively.}
\label{tab-3}
\setlength{\tabcolsep}{5mm}
\begin{tabular}{c|cc}
\hline
\hline
\diagbox [width=6em,trim=l]{$\ket{\psi(0)}$}{$t$} & $T$ & $2T$ \\
\hline
 $\ket{gss}$ & 0.9901 & 0.9655 \\
  $\ket{sgg}$ & 0.9812 & 0.9266 \\
 \hline
 $\ket{g\mathcal{T}_{-}s}$ & 0.9905 & 0.9662 \\
  $\ket{s\mathcal{T}_{+}g}$ & 0.9672 & 0.8717 \\
 \hline
 $\ket{g\mathcal{T}_{-}\mathcal{T}_{-}}$ & 0.9905 & 0.9662 \\
  $\ket{s\mathcal{T}_{+}\mathcal{T}_{+}}$ & 0.9489 & 0.8059 \\
   $\ket{gs\mathcal{Q}_{-\frac{3}{2}}}$ & 0.9908 & 0.9662 \\
  $\ket{sg\mathcal{Q}_{\frac{3}{2}}}$ & 0.9489 & 0.8059 \\
\hline
\hline
\end{tabular}
\end{table}

\begin{figure}
    \centering
    \includegraphics[width=0.48\textwidth]{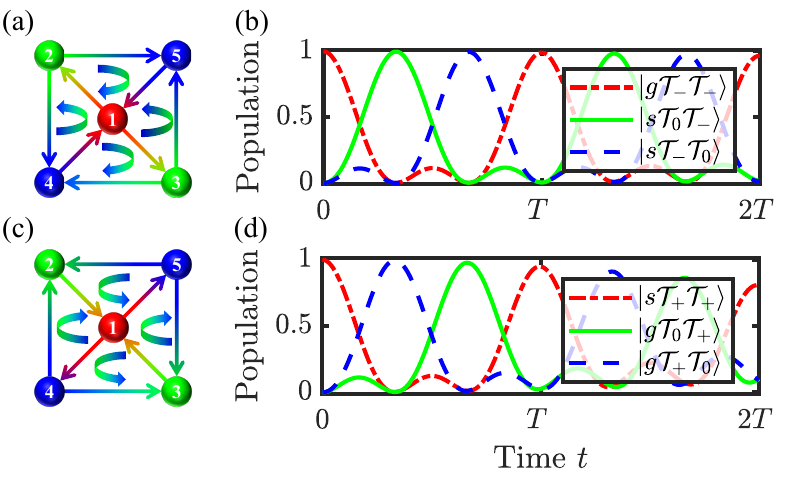}
    \caption{The schematic and population diagrams illustrate the evolution of chirality versus time generated from the states  $\ket{g\mathcal{T}_{-}\mathcal{T}_{-}}$ and $\ket{s\mathcal{T}_{+}\mathcal{T}_{+}}$, respectively. The dynamics is governed by the Hamiltonian in Eq. (\ref{eq_06}).}
    \label{fig06}
\end{figure}

\begin{figure}
    \centering
    \includegraphics[width=0.48\textwidth]{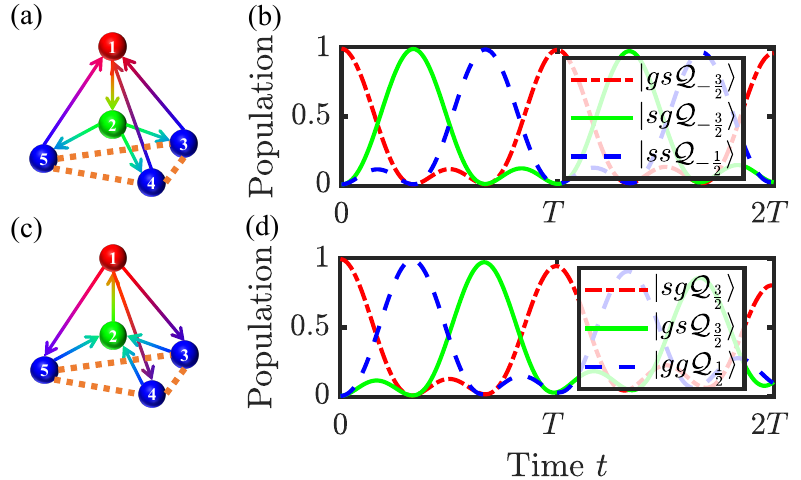}
    \caption{The schematic and population diagrams illustrate the evolution of chirality versus time generated from the states   $\ket{gs\mathcal{Q}_{-\frac{3}{2}}}$ and $\ket{sg\mathcal{Q}_{\frac{3}{2}}}$, respectively. The dynamics is governed by the Hamiltonian in Eq. (\ref{eq_06}).}
    \label{fig07}
\end{figure}

\begin{figure}
    \centering
\includegraphics[width=0.48\textwidth]{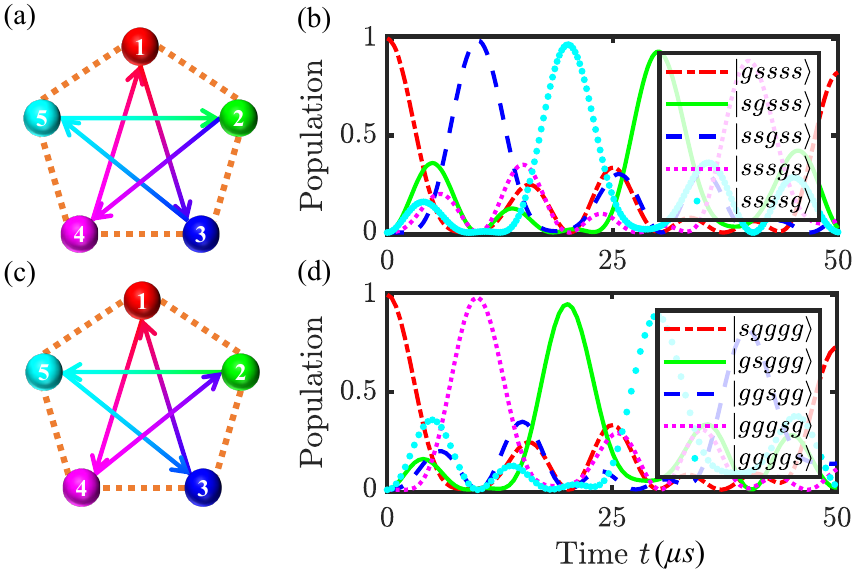}
    \caption{The schematic and population diagrams illustrate the evolution of chirality versus time generated from the states $\ket{gssss}$ and $\ket{sgggg}$, respectively. The dynamics is governed by the Hamiltonian in Eq. (\ref{eq_06}).}
    \label{fig08}
\end{figure}

For the case of five atoms, three distinct phase distributions have been selected in Sec \ref{SecII}. The first one closely resembles the previous distribution, and the corresponding evolutionary outcomes are illustrated in Fig. \ref{fig06}. It can be perceived that the state $\ket{g}$ traverses the sites in a counterclockwise manner. In a word, the site 1 advances to the superposition of sites 2 and 3, then progresses to the superposition of sites 4 and 5, and ultimately goes back to the site 1. For another scenario, the state $\ket{s}$ follows a opposite trajectory. It commences from the site 1, then passes through the superposition of sites 4 and 5, and transfers to the superposition of sites 2 and 3, eventually returns to site 1. As mentioned earlier, the second type of phase modulation follows a similar principle. It can be clearly seen from Fig. \ref{fig07}. When the system is initialized to the state $\ket{gs\mathcal{Q}_{-\frac{3}{2}}}$, the chiral current evolves a specific path: $1 \rightarrow 2 \rightarrow \{3, 4, 5\} \rightarrow 1 \cdots$. In another situation, the direction of evolution changes as $1 \rightarrow \{3, 4, 5\} \rightarrow 2 \rightarrow 1 \cdots$.
The above are the numerical simulation results for the three-state chiral current, where one clearly shows that the chirality emerges from the $2\pi/3$ phase shift between the three different populations oscillations. Moreover, the maximal value for all populations approaches to unit, indicating a perfect state transfer. Consequently, summarizing the results from the analysis above, our scheme is reasonable to open up exciting possibility for arbitrary atom numbers $N$, and that the transfer period remains independent of the choice of $N$ exactly.
\begin{table}[b]
    \centering
    \caption{The fidelity of chiral current with the initial state $\ket{s\mathcal{T}_{+}g}$ versus $\nu/\frac{\Omega_{k} g}{|\Delta_{k}|}$, such as $\mathcal{F}_{\ket{s\mathcal{T}_{+}g}} = |\langle \tilde{\psi}(t)| \psi(t) \rangle|^{2}$, where $\ket{\tilde{\psi}(t)}$ and $\ket{\psi(t)}$ are the evolution states of the full Hamiltonian in Eq. (\ref{eq_06}) and the effective Hamiltonian in Eq. (\ref{eq_17}) at time $t=T$, respectively.}
    \label{tab-4}
    \setlength{\tabcolsep}{3.5mm}
    \begin{tabular}{cccccc}
    \hline
    \hline
    $\nu/\frac{\Omega_{k} g}{|\Delta_{k}|}$ & 10 & 15 & 20 & 25 \\
    \hline
    Fidelity & 0.8032 & 0.9087 & 0.9486 & 0.9667 \\
    \hline
    \hline
    \end{tabular}
\end{table}
Nevertheless, the third kind exhibits a consistent phase difference between adjacent lattice sites in Fig. \ref{fig08}. The dynamics are driven by the Hamiltonian in Eq. (\ref{eq_06}), and the state $\ket{g}$ traverses the sites in the direction of $1 \rightarrow 3 \rightarrow 5 \rightarrow 2 \rightarrow 4 \rightarrow 1 \cdots$. Significantly, the maximum population approaches 1, which explains for the coupling strength between the nearest-neighbor and next-nearest-neighbor items not being absolutely proportional. Accordingly, combined with the other initial state, the propagation direction of the chiral current will be reversed.

Our finding reveals that the spin dynamics driven by the $z$-component of DMI exhibits a chiral evolution phenomenon. As a result, multiple instances of chiral motions are realized within various atomic configurations. Particularly, reversing the initial state of all sites, the chiral evolution has an opposite direction, yet this does not affect the time for completing a full cycle.

We further illustrate the trend of the fidelity versus time for chiral current across various sites $N$ and distinct initial states in Table \ref{tab-3}. It is evident from the overall pattern that the fidelity error of the chiral current rises as the number of sites $N$ rises. This is attributed to the fact that a larger $N$ entails a greater number of intricate sites, resulting in the chiral state transfer occurring within numerous smaller internal triangles. What's more, it is observable that the fidelity error for the chiral evolution of the states with $\ket{s}$ injected into one of the sites  exceeds that of the conditions with $\ket{g}$. Additionally, the validity of the Hamiltonian diminishes gradually over time, and so does the fidelity. In Table \ref{tab-4}, we present how the fidelity of the chiral current changes in relation to the parameter $\nu/\frac{\Omega_{k} g}{|\Delta_{k}|}$ by considering the initial state $\ket{s\mathcal{T}_{+}g}$ as an example. With an increase in the value of $\nu/\frac{\Omega_{k} g}{|\Delta_{k}|}$, the high-frequency approximation becomes increasingly precise, thereby bringing a more accurate effective Hamiltonian in Eq. (\ref{eq_11}).

\section{Conclusion}
\label{SecV}

To summarize, we simulate DMI and chiral current in atomic manifolds via periodic modulation. Utilizing a three-level system in a $\Lambda$ configuration, we derive an effective Hamiltonian for the ground-state subspace via adiabatic elimination, greatly mitigating the impact of decoherence. By simultaneously modulating the driving frequencies and initial phases, we achieve various configurations of DMI for different ground-state chiral dynamics, including three-state and multi-state chiral current. Additionally, we also obtain other components of DMI using the toggling frame. The numerical simulation results further confirm that the proposed method is capable to generate the ground-state chiral current. Our proposal of three-state chiral current can be extended to arbitrary atom numbers, and it is possible to realize the perfect multi-state chiral current by considering more tunable interaction between further neighbors.

\section*{Acknowledgments}
The authors would like to thank X. Q. Shao for discussion on this work. This work is supported by the Natural Science Foundation of Jilin Province (Grant No. JJKH20190279KJ), and the Key Lab of Advanced Optical Manufacturing Technologies of Jiangsu Province, Soochow University (Grant No. KJS2336). C.W. is supported by the National Research Foundation, Singapore, and A*STAR under its Quantum Engineering Programme (NRF2021-QEP2-02-P03).

\appendix
\renewcommand{\appendixname}{APPENDIX}

\section{THE EFFECTIVE HAMILTONIAN IN THE SLOW SUBSPACE}\label{appendixA}

Here, we introduce the projection operator technique used in this paper. First, we define the orthogonal projection operator of the atomic subspace as $\hat{P}=\sum_{k}(\hat{\sigma}_{k}^{gg}+\hat{\sigma}_{k}^{ss}), \hat{Q}=\sum_{k}\hat{\sigma}_{k}^{ee}$. Thus, in the Schr${\rm \ddot{o}}$dinger equation, we can project out the slow and fast variational subspaces as follows:
\begin{subequations}\label{eq_05}
  \begin{align}
   i \frac{d \hat{P}\ket{\Psi}}{d t}&=\hat{P} \hat{\tilde{H}}(t) \hat{P}\ket{\Psi}+\hat{P} \hat{\tilde{H}}(t) \hat{Q}\ket{\Psi}, \label{eq_05a} \\
   i \frac{d \hat{Q}\ket{\Psi}}{d t}&=\hat{Q} \hat{\tilde{H}}(t) \hat{P}\ket{\Psi}+\hat{Q} \hat{\tilde{H}}(t) \hat{Q}\ket{\Psi}. \label{eq_05b}
  \end{align}
\end{subequations}
Taking both sides of the equation in Eq. (\ref{eq_05b}) simultaneously with $[\hat{Q}\hat{\tilde{H}}(t)\hat{Q}]^{-1}$, we can derive
\begin{equation}\label{A1}
    \left[\hat{Q}\hat{\tilde{H}}(t)\hat{Q}\right]^{-1}i\frac{d\hat{Q}\ket{\Psi}}{d t}=\left[\hat{Q}\hat{\tilde{H}}(t)\hat{Q}\right]^{-1}\hat{Q}\hat{\tilde{H}}(t)\hat{P} \ket{\Psi}+ \ket{\Psi}. 
\end{equation}
Acting the projection operator $\hat{Q}$ on both sides of Eq. (\ref{A1}), we obtain
\begin{equation}\label{A2}
\small
    \hat{Q}\ket{\Psi}  = \hat{Q}\left[ \hat{Q}\hat{\tilde{H}}(t)\hat{Q} \right]^{ - 1}i\frac{d\hat{Q}\ket{\Psi}}{dt} - \hat{Q}\left[\hat {Q}\hat{\tilde{H}}(t)\hat{Q} \right]^{ - 1}\hat{Q}\hat{\tilde{H}}(t)\hat{P}\ket{\Psi}. 
\end{equation}
Given the condition of large detuning and an initial state with no population in the excited state, the term $\hat{Q}[\hat{Q}\hat{\tilde{H}}(t)\hat{Q}]^{-1}id\hat{Q}\ket{\Psi}/dt$ can be safely omitted. Bring back the integrated $\hat{Q}\ket{\Psi}\approx-\hat{Q}[\hat{Q}\hat{\tilde{H}}(t)\hat{Q}]^{-1}\hat{Q}\hat{\tilde{H}}(t)\hat{P}\ket{\Psi}$ to Eq. (\ref{eq_05a}), we can receive
\begin{equation}\label{A3}
\small
   i\frac{d\hat{P}\ket{\Psi}}{d t}\approx\hat{P}\hat{\tilde{H}}(t)\hat{P}\ket{\Psi}-\hat{P}\hat{\tilde{H}}(t)\hat{Q}\left[\hat{Q}\hat{\tilde{H}}(t)\hat{Q}\right]^{-1}\hat{Q}\hat{\tilde{H}}(t)\hat{P}\ket{\Psi}. 
\end{equation}
Considering the relation $\hat{P}^{2} = \hat{P}$ , we obtain
\begin{equation}\label{A4}
    i\frac{d\hat{P}\ket{\Psi}}{dt}\ \approx\hat{H}^{\prime}(t)\hat{P}\ket{\Psi},
\end{equation}
with 
\begin{equation}
    \hat{H}^{\prime}(t)=\hat{P}\hat{\tilde{H}}(t)\hat{P}-\hat{P}\hat{\tilde{H}}(t)\hat{Q}\left[\hat{Q}\hat{\tilde{H}}(t)\hat{Q}\right]^{-1}\hat{Q}\hat{\tilde{H}}(t)\hat{P}.
\end{equation}
Hence, the effective Hamiltonian expressed in Eq. (\ref{eq_06}) is derived within the ground-state subspace.

\bibliography{ref}

\vspace{8pt}

\end{document}